\documentclass[10pt,letterpaper]{article}
\usepackage{opex3}
\usepackage{multirow}

\begin{document}

\title{Measurement of focusing properties for high numerical aperture optics using an automated submicron beamprofiler}

\author{J. J. Chapman, B. G. Norton, E. W. Streed and D. Kielpinski}

\address{Centre for Quantum Dynamics, Griffith University, Queensland, Brisbane 4111 Australia}

\email{justin.chapman@student.gu.edu.au} 


\begin{abstract} The focusing properties of three aspheric lenses with numerical aperture (NA) between 0.53 and 0.68 were directly measured using an interferometrically referenced scanning knife-edge beam profiler with sub-micron resolution.  The results obtained for two of the three lenses tested were in agreement with paraxial gaussian beam theory. It was also found that the highest NA aspheric lens which was designed for 830nm was not diffraction limited at 633nm. This process was automated using motorized translation stages and provides a direct method for testing the design specifications of high numerical aperture optics.  \end{abstract}



\section{Introduction}

There are a number of processes which rely on tightly focusing light for which the shape and size of the spot must be known to achieve the desired outcome. For example, optical memory is now a common method for data storage.  As advances in other aspects of these systems are made (such as increased resolution of optical pickup [\ref{bib:opstorage}]) the optics must focus higher frequency laser light tightly and reliably.
Optical trapping by tightly focused lasers has been widely used for the manipulation of sub-micron sized particles.  The force in these traps is proportional to the gradient of intensity, and is therefore controlled by the quality and spot size of the focusing lens [\ref{bib:ref6},\ref{bib:optrap}].  Improving the signal-to-noise ratio in fluorescence spectroscopy of single molecules is achieved by reducing the detection volume determined by the spot size which is directly related to the numerical aperture of the focusing objective.  To achieve maximum collection efficiency it is therefore very important that the spot size and quality of the lens is as expected [\ref{bib:singmol1},\ref{bib:singmol2}].
Advanced optical imaging methods such as stimulated emission depletion microscopy [\ref{bib:STED}] or 4$\pi$ microscopy [\ref{bib:4pi}] need lenses with the highest numerical aperture possible to achieve the highest imaging resolution and fluorescence collection efficiency. All these applications require precise knowledge of the lens's focal spot to correctly predict and evaluate experimental results [\ref{bib:eval1},\ref{bib:eval2}]. It is therefore necessary to first be able to accurately characterize the optic's focusing ability.

High numerical aperture (NA) optics are currently characterized using three common test methods [\ref{bib:opshoptest}]; 3d profilometers, null tests and the Hartmann test .  These methods measure surface geometry of optics and use that information to reconstruct the element's focusing properties.  It is more reliable to characterize the optic's performance at the focus directly as this is the experimentally most relevant position.
While there are many beam profiling units commercially available for this purpose, none have adequate resolution for measuring submicron spots.  The knife-edge scanning technique is a well known process which has the ability to perform submicron waist measurements. This process finds the beam waist by scanning a razor through the beam and measuring the corresponding change in transmitted power. This has been previously been used to characterize submicron spots [\ref{bib:rotblade}-\ref{bib:piezo}], but these experiments lacked the automation required for efficient and rapid testing.  We used motorized translation stages to automate the test process, with accurate velocity calibration provided by an interferometer. Input beam size and spot size were measured for a number of aspheric lenses in order to demonstrate the apparatus.

\section{Apparatus}
\begin{figure}
\centering\includegraphics[width=50mm]{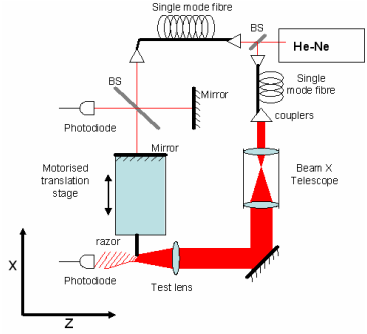}
\caption{Beam profiling apparatus-translation stage scans the razor through the beam while acting as one of the interferometer arms providing accurate distance calibration for the waist measurement of the focused beam}\label{fig:apparatus}
\end{figure}

 In order to measure the optics' focusing ability and investigate the spot size directly, an automated beam profiler with submicron resolution was constructed.  The apparatus is shown in Fig.\ \ref{fig:apparatus}. A razor is attached to one side of the translation stage and a mirror is attached to the opposite side.  In this way the stage acts simultaneously as the scanning element and the reflector for one of the interferometer arms providing an accurate velocity calibration in the direction perpendicular (x-axis Fig.\ \ref{fig:apparatus}) to the beam.
632.8nm light from a helium-neon laser was used in the interferometer and for testing the lens.  Two precision motorized translation stages with resolution of about 50nm were used to automate the beam profiling process. The power in the beam is measured real-time while the razor is cutting the beam through the x-axis in Fig.\ \ref{fig:apparatus} , and the interferometer calibrates the razor's velocity. By observing the interference fringes it was discovered that the motors do not move at their nominal velocity, implying the interferometer is critical to obtain an accurate result.  In contrast, errors in stage movement parallel to the beam (the z-axis in Fig.\ \ref{fig:apparatus}) were not found to limit the measurements.

The laser light was delivered to two single mode fibers to provide increased flexibility and reliability of the system.
One fiber output provides light to the interferometer. The other output is collimated, sent through an expansion telescope of variable magnification and coupled into the test lens. The lens position and the beam's angle of incidence is controlled by an xyz translation stage and mirror.  Proper alignment of the beam through the lens is achieved by measuring the dependence of the beam waist on each of the four alignment variables - horizontal and vertical angle of incidence, and horizontal and vertical position of lens, then setting each variable to minimize the observed waist size.  This procedure is iterated until convergence.  A gimbal mount was used for the mirror immediately before the test lens to ensure the horizontal and vertical changes of angle are decoupled. Once the lens was properly aligned the knife-edge scanning method was used to obtain waist measurements along the length of the focused beam.
To ensure that the roughness of the blade did not affect the measured spot sizes we obtained an SEM image of the razor edge. The error in the waist measurement can be assumed insignificant for spatial variations outside the scale of $\frac{d}{10}\geq\delta$x$\geq$10d where d is the 1/$e^2$ diameter of the beam and $\delta$x is the size of the variation.  The RMS roughness over this range was 0.035$\mu$m, negligible compared to other errors.

\section{Results and Discussion}
\begin{figure}
\centering\includegraphics[width=7cm]{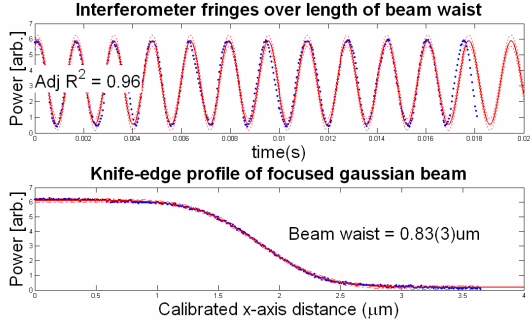}
\caption{Typical interferometer (upper panel) and knife-edge (lower panel) data for a single knife-edge cut. Dots: Data, Solid line: fit to sinusoid (upper panel) and Gaussian power distribution (lower panel)}\label{fig:raw}
\end{figure}

Paraxial Gaussian theory predicts that a beam of $1/e^2$ radius $w_1$ incident on a lens with focal length f will focus to a spot size $w_0$ under the relation given by Eq.\ \ref{eq:w0} [\ref{bib:siegman}], where the spot size $w_0$ is the $1/e^2$ radius of the beam at the focus.
\begin{equation}
w_0w_1\approx\frac{f\lambda}{\pi }
\label{eq:w0}
\end{equation}
This equation was used to calculate the theoretical spot sizes for each combination of lens and input beam size.  The raw data obtained from the apparatus consists of an error function representing the power in the beam, and sinusoidal interference fringes from the interferometer as shown in Fig.\ \ref{fig:raw}.   The interference fringes are fit with a sinusoidal function of the form $Asin(a_0+a_1t+a_2t^2+a_3t^3)+B$ where \textit{$A$}, \textit{$B$}, \textit{$a_1$}, \textit{$a_2$}, and \textit{$a_3$} are fit coefficients and \textit{t} is time.  It was found that adding higher order terms in terms did not significantly improve the fit.  The change in power of the beam as the razor was scanned across was measured over time. The time axis was then converted to a calibrated distance scale for every scan. The x-axis distance for each set of knife-edge data was re-fit with its calibration data allowing the waist of the beam to be known.

Approximately 20 of these scans are recorded at intervals along the length of the beam centered around the focal point.
The waist measurements from each of these scans was tabulated and fit with Eq.\ \ref{eq:M2}.  This equation gives the spot size and Rayleigh range of the beam.
\begin{equation}
w(z)=w_0\sqrt{1+[z/z_R]^2}
\label{eq:M2}
\end{equation}

The Rayleigh range of an ideal gaussian beam with the same waist $w_0$ is calculated using the equation Eq.\ \ref{eq:Rayleigh}.
\begin{equation}
z_R=\frac{\pi w_0^2}{\lambda}
\label{eq:Rayleigh}
\end{equation}
The ratio of the ideal and measured Rayleigh ranges $z_{R}^{ideal}/z_{R}^{meas}$ gives the $M^2$ value for that particular beam expansion.

We tested three aspheric lenses from Kodak and Lightpath, their properties are summarized in table 1.
\begin{table}
\begin{minipage}[b]{\textwidth}
\centering
\begin{tabular}[b]{c c c c c}
\hline
\multicolumn{5}{c}{\bf{Table 1. Design specifications of aspheric lenses}} \\
\hline

Lens&NA&Focal length&Design wavelength&Clear aperture\\
    &  & (mm)       &  (nm)           & (mm)\\
\hline
Kodak A390	    &0.53	&4.6	&655	&4.89\\
LightPath 350330	&0.68	&3.1	&830	&5\\
LightPath 352671	&0.6	&4.02	&408	&4.8\\ \hline
\end{tabular}
\caption{Properties of aspheric lenses used for testing.}
\label{tab:lenses}
\end{minipage}
\end{table}
Waist measurements for the collimated output from the single-mode fiber were recorded using a commercial CCD beam profiler over 1.2m. The spot size and $M^2$ of the beam were determined to be 350(30)um and 1.02(1) respectively, so the input beam is well approximated by a pure Gaussian beam. The beam expansion telescopes introduced a maximum beam divergence of 90$\mu$rad.  This should not affect the $M^2$ but introduces a maximum error of 0.1$\mu$m to the measured spot size. The telescope optics were large enough to ensure a negligible contribution to diffraction of the beam at the largest beam expansion.

\begin{figure}
\centering\includegraphics[width=8cm]{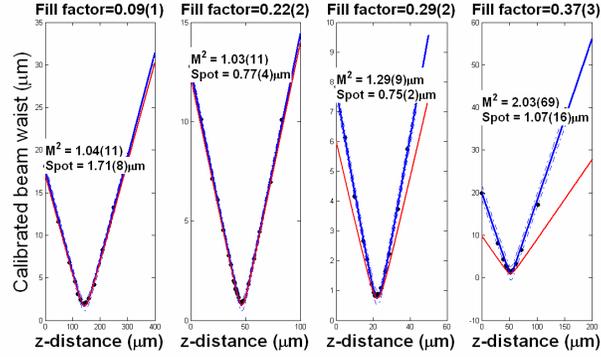}
\caption{$M^2$ and spot dependence on fill-factor for Kodak A390 lens.  Dots: Data, Red solid line: Ideal gaussian and Blue solid line: measured beam.}\label{fig:A390}
\end{figure}

The intensity ripples in the near field caused by diffraction are approx 1\% of the amplitude at the condition $2\textit{a} = 4.6w_0$ where \textit{a} is the radius of the clear aperture and \textit{$w_0$} is the waist of the input beam [\ref{bib:siegman}]. For larger input beam sizes diffraction effects become significant and have the effect of increasing the spot size, divergence and $M^2$ of the beam.  To facilitate comparisons of the different lenses, we define the \textit{fill-factor} as the ratio of the input beam waist and clear aperture, so that the 1\% amplitude criterion corresponds to a fill-factor of 0.22 in all cases.  The error in the fill-factors in Fig.\ \ref{fig:A390} arises from the uncertainty in the input beam waist. The Kodak A390 and LightPath 352671 lenses both showed an increase in the $M^2$ at this condition. The results for the Kodak A390 lens are shown in Fig.\ \ref{fig:A390}.  The figure shows the divergence for the actual beam in blue and the divergence for an ideal gaussian beam in red.

The LightPath 350330 however showed an increase in $M^2$ at 0.14(1) fill-factor.  Only four data points were taken for this lens since its performance was not diffraction limited as demonstrated by the rapid increase in $M^2$. The LightPath 350330 aspheric lens is the highest numerical aperture lens readily available commercially, and also has the largest clear aperture out of the three lenses tested.  It is designed to operate at 830nm and was tested at 632.8nm, which could explain the deviation from diffraction limited performance.

\begin{figure}
\centering\includegraphics[width=9cm]{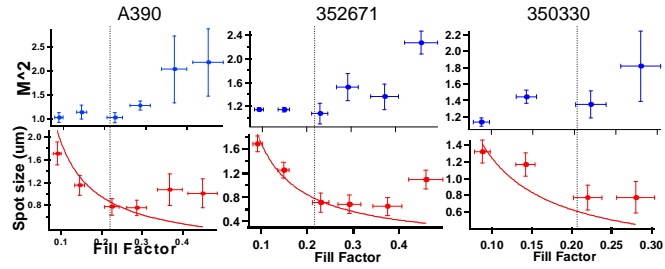}
\caption{Spot size and $M^2$ as a function of fill-factor for each lens.  The condition of 1\% diffraction effects is represented by the dotted line. Dots: Data, Solid line: Theoretical spot size (lower panel).}
\label{fig:collaboration}
\end{figure}

These results are summarized in Fig.\ \ref{fig:collaboration}, which demonstrate the dependence of spot size and $M^2$ on input beam size. The error in the fill-factor in Fig.\ \ref{fig:collaboration} is also due to the uncertainty in input beam waist, as in Fig.\ \ref{fig:A390}.

\section{Conclusion}
An automated method for testing the focusing properties of high numerical aperture optics with submicron resolution was demonstrated.  Test results for aspheric lenses  of NA up to 0.68 were in agreement with the limits of paraxial gaussian beam theory with the inclusion of clipping effects for input beam sizes that overfilled the lens aperture. It was further determined that one of the aspheric lenses was not diffraction limited, possibly because the lens was not tested at its design wavelength.

\section{Acknowledgements}
This work was supported by the US Air Force Office of Scientific
Research under contract FA4869-06-1-0045, Prof. Howard Wiseman's
Australian Research Council Federation Fellowship grant FF0458313, and
the Australian Research Council Discovery Project grant DP0773354.
\end{document}